\documentstyle[aps,prl,preprint]{revtex}
\draft
\begin{document}
\title{Extension of a Spectral Bounding Method to 
Complex Rotated Hamiltonians, with Application to 
$p^2-ix^3$ 
}
\author{C. R. Handy and Xiao Qian Wang}
\address{Department of Physics \& Center for Theoretical Studies of 
Physical Systems, Clark Atlanta University, 
Atlanta, Georgia 30314}
\date{Received \today}
\maketitle
\begin{abstract}
We show that a recently developed method for generating bounds for the 
discrete energy states of the non-hermitian $-ix^3$ potential (Handy 2001) is 
applicable to complex rotated versions of the Hamiltonian. This has
important implications for extension of the method in the analysis of 
resonant states,  Regge poles, and general bound states in the complex 
plane (Bender and Boettcher (1998)).

\end{abstract}
\vfil\break
\section{Introduction}

In a recent work, Handy (2001) presented a novel quantization formalism
for generating converging bounds to the (complex) discrete spectra 
of non-hermitian, one dimensional, potentials.
 The first part of this analysis makes use of the fact that
the modulus squared of the wavefunction, $S(x) \equiv 
|\Psi(x)|^2$, for $x \in \Re$, and any (complex) potential function, $V(x)$,
satisfies a fourth order, linear  differential equation:
\begin{eqnarray}
-{1\over {V_I-E_I}}S^{(4)} - \Big({1\over{V_I-E_I}}\Big)'S^{(3)}
+ 4 \Big( {{V_R-E_R}\over {V_I-E_I}} \Big) S^{(2)} \cr
+ \Big( 4\Big({{ V_R}\over { V_I-E_I}}\Big)'+ 2\Big({{{V_R}'}\over
{ V_I-E_I}}\Big ) 
-4 \Big({{E_R}\over{ V_I-E_I}}\Big)' \Big) S^{(1)} 
+ \Big( 4  (V_I-E_I) + 2\Big({{{V_R}'}\over { V_I-E_I}}\Big)'\Big) S = 0,
\end{eqnarray}
where $S^{(n)}(x) \equiv \partial_x^nS(x)$, and $E = E_R+iE_I$, etc.

 The
bounded ($L^2$) and nonnegative solutions of this differential equation
 uniquely correspond to
the physical states. For the case of rational fraction potentials,
there is an associated (recursive) moment equation,
 whose solutions are parameterized by the energy, $E$. Constraining these
through the appropriate Moment Problem conditions for nonnegative 
functions  (Shohat
and Tamarkin (1963)), generates rapidly converging 
bounds on the discrete state energies.
This entire procedure is referred to as the
Eigenvalue Moment Method (EMM) and was originally
 developed by Handy and Bessis (1985)
and Handy et al (1988a,b). An important theoretical and algorithmic component
is the use of linear programming (Chvatal (1983)).

The EMM procedure is not solely a numerical implementation of some very 
important mathematical theorems.
 It is possible to implement it algebraically, deriving
bounding formulas for the eigenenergies (Handy and Bessis (1985)). Furthermore,
it does not depend on the hermitian or non-hermitian structure of the 
Hamiltonian. What is important is that the desired solutions be the unique
nonnegative and bounded solutions of the differential system being investigated.
In this sense, it provides an important example of how positivity can be
used as a quantization procedure.

Many important problems can be analyzed through the application of EMM on
Eq.(1), and its generalization as outlined in Sec. II. This 
includes generating bounds to Regge poles (which first motivated Handy's
formalism),
 to be presented in a
 forthcoming work (Handy and Msezane (2001)).
Other problems include the calculation of resonant states as well as
 bound states in the complex plane as discussed by
Bender and Boettcher (1998), and elaborated upon by others 
cited in the more recent work of Mezincescu (2001).

However, the realization of these depends on understanding the complex rotation
extension of Handy's original formulation. Indeed, such concerns
introduce a new class of EMM problems not encountered before, and thereby 
motivate the present work. 

Our basic objective  is to extend the $S(x)$-EMM formulation
to complex rotated transformations of the Hamiltonian $p^2 -ix^3$. We 
are particularly interested in the effectiveness of
such an analysis. Indeed, once the correct moment equation is developed,
one finds that the EMM approach works very well in this case, and yields
the anticipated result, as detailed in Sec. III. 

In the following sections we present a more general 
derivation of the $S(x)$ equation, and
apply the resulting formalism to the rotated $-ix^3$ Hamiltonian.

\vfil\break
\section { Generalized $S$-Equation}

Consider the (normalized) Schrodinger equation

\begin{equation}
-\partial_x^2\Psi(x) + {\cal V}(x) \Psi(x) = E \Psi(x),
\end{equation}
for complex energy, $E$, and complex potential, ${\cal V}(x)$.
Assume that the (complex) bound state, $\Psi(x)$, lies within the 
complex-$x$ plane, along some infinite contour, ${\cal C}$. Let
$x(\xi)$ define a differentiable map from a subset of the real axis
to the entire complex contour:

\begin{equation}
x(\xi): \xi \in \Re \rightarrow {\cal C}.
\end{equation}
The transformed Schrodinger equation is

\begin{equation}
-\Big( D(\xi)\partial_\xi\Big )^2\Psi(\xi) + V(\xi)\Psi(\xi) = E \Psi(\xi),
\end{equation}
where $D(\xi) \equiv (\partial_\xi x)^{-1}$, and $V(\xi) \equiv {\cal V}(x(\xi))$. Alternatively, we may rewrite the above as

\begin{equation}
H_\xi\Psi(\xi) \equiv  A(\xi)\partial_\xi^2\Psi(\xi) + B(\xi) \partial_\xi\Psi(\xi) +
C(\xi) \Psi(\xi) = 0,
\end{equation}
where $A(\xi) \equiv - (D(\xi))^2$, $B(\xi) = -{1\over 2} \partial_\xi (D(\xi))^2$, and 
$C(\xi) = {V(\xi)-E}$.
In the previous work by Handy (2001), he showed how such an expression, without 
the $\partial_\xi\Psi$ term, leads to a fourth order equation for $S(\xi) = 
|\Psi(\xi)|^2$. We broaden this derivation to include the above general case.

Define the following four functions:

\begin{equation}
\Sigma_1(\xi) \equiv \Psi^*(\xi) H_\xi\Psi(\xi) + c.c. = 0,
\end{equation}
\begin{equation}
\Sigma_2(\xi) \equiv \Psi'^*(\xi) H_\xi\Psi(\xi) + c.c. = 0,
\end{equation}
\begin{equation} 
\Delta_1(\xi) \equiv \Psi^*(\xi) H_\xi\Psi(\xi) - c.c. = 0.
\end{equation}
and
\begin{equation}
\Delta_2(\xi) \equiv \Psi'^*(\xi) H_\xi\Psi(\xi) - c.c. = 0.
\end{equation}

Define 
\begin{equation}
S(\xi) = \Psi^*(\xi) \Psi(\xi),
\end{equation}
\begin{equation}
P(\xi) =  \Psi'^*(\xi) \Psi'(\xi),
\end{equation}
\begin{equation}
J(\xi) = {{\Psi(\xi)\partial_\xi\Psi^*(\xi) - \Psi^*(\xi)\partial_\xi\Psi(\xi)}
\over {2i}},
\end{equation}
and
\begin{equation}
T(\xi) = {{\partial_\xi\Psi(\xi)\partial_\xi^2\Psi^*(\xi) - \partial_\xi\Psi^*(\xi)\partial_\xi^2\Psi(\xi)}
\over {2i}}.
\end{equation}
These correspond to important physical quantities. The first two are nonnegative
functions corresponding to the probability density and the ``momentum density",
while $J(\xi)$ is the probability flux.

We then have (i.e. $A = A_R+ i A_I$, etc.):
\begin{equation}
\Sigma_1(\xi) = (S''(\xi) - 2 P(\xi))A_R(\xi)  + S'(\xi) B_R(\xi) + 2 S(\xi) C_R(\xi) + 2 (B_I(\xi) +  A_I(\xi) \partial_\xi)J(\xi)   = 0,
\end{equation}
\begin{equation}
{{\Delta_1(\xi)}\over {i}} =  \Big(  S''(\xi) - 2 P(\xi)\Big) A_I(\xi) + S'(\xi) B_I(\xi) + 2S(\xi) C_I(\xi) - 2( B_R(\xi) + A_R(\xi)\partial_\xi))J(\xi)  = 0,
\end{equation}
\begin{equation}
\Sigma_2(\xi) = P'(\xi) A_R(\xi) + 2 T(\xi) A_I(\xi) + 2P(\xi) B_R(\xi) + S'(\xi) C_R(\xi) - 2 J(\xi) C_I(\xi) = 0,
\end{equation}
and
\begin{equation}
{{\Delta_2(\xi)}\over{i}} = P'(\xi) A_I(\xi) - 2T(\xi) A_R(\xi)  + 
2P(\xi) B_I(\xi)  + S'(\xi) C_I(\xi) + 2J(\xi) C_R(\xi)  = 0.
\end{equation}

For the case of the ordinary (real potential) Schrodinger equation where $x = \xi$, and $A_I = 0$,$B = 0$, and $C_I = 0$, we can combine Eq.(14) (divided by 
$A_R$) and Eq.(16) to generate a third
order differential equation for $S$ (Handy (1987a,b), Handy et al (1988c)).

If $A_I = 0$ and $B = 0$, the case considered by Handy (2001), then 
Eq.(14) (divided by $A_R$) yields an equation for $P$. Differentiating this
relation, and inserting Eq. (16) defines a third order equation in $S$, which
also depends on $J$. We can elliminate $J$ by differentiating (after dividing 
out ${{C_I}\over {A_R}}$) one more time,
and substituting Eq.(15). The resulting expression corresponds to that 
in Eq.(1), assuming $A_R = 1$.

If $A_I = 0$ and $C_I \neq 0$, or $A_R = 0$ and $C_I \neq 0$, we 
can determine $J$ from Eq.(16) or Eq.(17), respectively, and procede with
the analysis discussed below.

If both $A_{R,I} \neq 0$, multiplying Eq.(16) and Eq.(17) by $A_R(\xi)$ and $A_I(\xi)$,
respectively, and adding, yields an equation for $J$:

\begin{equation}
J(\xi) = {{P'(\xi) |A(\xi)|^2 + 2 P(\xi)Re(A(\xi)B^*(\xi)) + S'(\xi) Re(A(\xi)C^*(\xi))}\over{-2Im(A(\xi)C^*(\xi))}}.
\end{equation}

We can now substitute this expression, $J(\xi;P,P',S')$, into Eq.(14) and Eq.(15) yielding two coupled differential equations for $P$ and $S$. Instead of the 
formalism presented by Handy (2001), we can work with these two coupled equations, since they involve two nonnegative configurations (for the physical solutions). However, depending on the nature of the function  coefficients (i.e. $A(\xi), B(\xi), C(\xi)$) it may be expedient to reduce these to one fourth order,
linear differential equation for $S$. This is particularly the case  corresponding
to that discussed by Handy (2001).

For completeness, we note that the analytic extension of $S(\xi)$ onto the
complex-$\xi$ plane (which is of no immediate concern to us)
 corresponds to
\begin{equation}
S(\xi) = \Psi^*(\xi^*)\Psi(\xi).
\end{equation}
That is, if we assume that $\Psi(\xi)$ is analytic at the origin (for simplicity), then $\Psi(\xi) = \sum_{j=0}^\infty c_j \xi^j$ and 
$\Psi^*(\xi^*) = \sum_{j=0}^\infty c_j^* \xi^j$ are convergent power series, 
defining an analytic product which is nonnegative along the real $\xi$-axis.

\vfil\break

\section{ The Complex Rotated $p^2+(ix)^3$ Hamiltonian}

Consider the non-hermitian potential problem

\begin{equation}
-\Psi''(x) -ix^3 \Psi(x) = E \Psi(x),
\end{equation}
first considered in the work by  Bender and Boettcher (1998). 
From their analysis, it is known that bound states exist along the real axis. 
They satisfy

\begin{equation}
E = {{\int dx P(x) -i \int dx  x^3 S(x)}\over {\int dx S(x)}}.
\end{equation}
If the physical solution conforms with the ${\cal P}{\cal T}$ invariant nature
of the Hamiltonian,
\begin{equation}
\Psi^*(-x) = \Psi(x),
\end{equation}
then $S(-x) = S(x)$, and the discrete state energy is real. Exhaustive 
investigations by many individuals supports this. These include the
works by (in addition to those cited)
Bender et al (1999), Bender et al (2000),
Bender and Wang (2001),  Caliceti (2000), Delabaere and Pham (1999), Delabaere and Trinh (2000),
Levai and Znojil (2000), Mezincescu (2000,2001),
Shin (2000), and Znojil (2000). The analysis of Handy(2001) has also been extended to the case of complex $E$, also supporting that the discrete states of the
$-ix^3 $ potential do not violate  ${\cal P}{\cal T}$ symmetry (Handy  
 (2001)). In accordance with all of this, we procede assuming that
the discrete spectrum is real.

Let $x \equiv e^{i\theta}|x|$.  Adopting Bender and
Boettcher's notation, it is known that the bound states
decrease exponentially fast in any asymptotic direction 
within the `wedges' defined by

\begin{equation}
|\theta - \theta_{L,R}| < {\pi\over 5},
\end{equation}
where $\theta_L = -\pi + {\pi\over {10}}$, and $\theta_R = -{\pi\over {10}}$.
The anti-Stokes lines (along which the asymptotic decrease is fastest) is 
centered at $\theta_{L,R}$. 

We first considered the general transformation

\begin{equation}
x(\xi) = e^{i\theta} \xi - {{e^{-i\theta}}\over \xi},
\end{equation}
for $0 < \xi < \infty$, and $\theta$ lying within the boundedness wedge(s). 
Such transformations would be relevant not only for the $-ix^3$ potential,
but all the other non-Hermitian potentials considered by Bender and Boettcher. However, the generation of a moment equation, following the formalism in Section II,
made this approach inefficient, due to the large order of the associated
moment equation relation.

Instead of this approach, the $-ix^3$ problem affords a particularly simpler
formulation, since the wedges intersect  the entire real axis. This is not
the case for all the other potentials (i.e. $(-ix)^n$, $n \geq 4$)
 considered by 
them. The generalization of the present approach can be made to these cases 
as well; however, we defer this to another communication, since it requires 
a completely different linear programming formalism.

Consistent with the formalism in Sec. II, if we take
\begin{equation}
x = e^{i\theta} \xi,
\end{equation}
where $\xi \in (-\infty,+\infty)$, then the bound state solution $\Psi(e^{i\theta} \xi)$ will be asymptotically zero (i.e. at $\xi = \pm \infty$) for 
\begin{equation}
-{\pi\over{10}} < \theta < {\pi \over {10}}.
\end{equation}
Clearly, this includes the real axis. However, for fixed 
$\theta \neq 0$, the closer
$e^{i\theta}\xi$ is to the anti-Stokes line along the positive real-$\xi$ axis
(i.e. $\theta \rightarrow \theta_R$) the farther is $e^{i\theta}\xi$ from 
its respective anti-Stokes line, for $\xi < 0$ (and conversely). Only along the
real axis ($\theta = 0$), will the exponential decay be equally balanced between both the positive and negative real axis.

The work by Handy(2001) introduces a very accurate bounding technique which was 
used to generate the first five low lying states. Our immediate interest is 
in testing the same EMM formalism with respect to complex rotations of the
system.

\subsection{ Moment Equation for the Complex Rotated $-ix^3$ Potential}

As noted, we will be solving the Schrodinger equation along the complex-$x$ 
infinite contour defined by $x = e^{i\theta}\xi$, for $-\infty < \xi < +\infty$. The resulting differential equation is

\begin{equation}
-\partial_\xi^2\Psi(\xi) -i e^{i5\theta}\xi^3\Psi(\xi) =  E e^{i2\theta}\Psi(\xi).
\end{equation}
We will assume $E$ to be real. The resulting, fourth order differential equation for $S(\xi)$ is then
\begin{eqnarray}
{1\over {\Lambda(\xi)}}S^{(4)}(\xi)
-{{3c_5\xi^2}\over {\Lambda(\xi)^2}} S^{(3)}(\xi)
+ {{4c_2E-4\xi^3s_5}\over {\Lambda(\xi)}} S^{(2)}(\xi)
-{{6\xi^2(2c_2c_5E + c_5s_5\xi^3 + 3Es_2s_5)}\over {\Lambda(\xi)^2}} 
 S^{(1)}(\xi) \cr
-{{4c_5^3 \xi^9 + 12 c_5^2 s_2 E \xi^6 +12 c_5 s_2^2E^2\xi^3 +4 E^3 s_2^3 
- 6c_5 s_5 \xi^4 +12 E s_2s_5\xi}\over {\Lambda(\xi)^2}} S^{(0)}(\xi) = 0,
\end{eqnarray}
where 
\begin{equation}
\Lambda(\xi) = c_5 \xi^3 + E s_2,
\end{equation}
$c_n \equiv cos(n\theta)$,  $s_n \equiv sin(n\theta)$, and $S^{(n)}(\xi) \equiv 
\partial_\xi^nS(\xi)$.

We know that all of the solutions to this equation are regular in the entire 
complex $\xi$ plane, despite the singular function coefficient, ${1\over \Lambda}$ (Handy(2001)). There is a real singularity at $\xi = b$, defined by
\begin{equation}
b = -\big({{E s_2}\over {c_5}}\big)^{1\over 3} \in \Re.
\end{equation}
In terms of the implicit $\theta$ dependence, $\lim_{\theta \rightarrow 0}b(\theta) = 0$.  

In all of the previous 
applications of the EMM approach, we have never considered problems with  function coefficients singular within the physical domain. The types of
singularities encountered have either been  close to the physical domain,
or at the boundaries. 

Despite the fact that the singularities in question
 do not affect the regular nature of
the solutions, they are of concern within the EMM formalism, as
discussed in the recent work by
Handy, Trallero-Giner, and Rodriguez (HTR, 2001).  Intuitively, let us represent
the above fourth order equation by ${\cal O}_\xi S(\xi)$. If one is to generate
a moment equation by multiplying both sides of this differential equation by
$\Lambda(\xi)^2$, one must make sure that the resulting moment equation does 
not admit solutions to the more general problem of ${\cal O}_\xi S(\xi) = {\cal D}(\xi-b)$, where ${\cal D}(\xi-b)$ is a distribution-like expression supported at
$b$. Such expressions disappear when multiplied by $\Lambda(\xi)$ (with a 
zero at $b$). If one
is not careful, the naive moment equation may not correspond to the desired
problem, ${\cal D} = 0$. That is, an inappropriate specification of the moment equation would not produce any converging bounds.
Such concerns are generally inconsequential, within the EMM framework, if
it is known that the desired solution also has a zero at $b$ (HTR, 2001). For the present
problem, this is not the case, and a suitable modification of the standard EMM
formalism is required.

In order to address this problem, one should first perform a translation change
of variables:
\begin{equation}
\chi = \xi - b.
\end{equation}
The ensuing fourth order differential equation will then involve function coefficients whose singularity is no more singular than ${1\over{\chi^2}}$. Specifically, let $\Lambda(\chi) = \chi \Upsilon(\chi)$, where $\Upsilon(\chi) = c_5
(3b^2 + 3b\chi + \chi^2) \neq 0$. Upon multiplying both sides of the translated 
fourth order equation (${\cal O}_\chi S(\chi) = 0$) by $\Upsilon(\chi)$ , 
one obtains an expression of the form

\begin{equation}
{{C_4(\chi)}\over {\chi}} S^{(4)}(\chi) + {{C_3(\chi)}\over {\chi^2}} S^{(3)}
(\chi) +{{C_2(\chi)}\over {\chi}} S^{(2)}(\chi) +  {{C_1(\chi)}\over {\chi^2}}
 S^{(1)}(\chi) + {{ C_0(\chi)}\over {\chi^2}} S(\chi) = 0,
\end{equation}
where all the $C_n(\chi)$ are polynomials in $\chi$.

Let us now multiply by $\chi^\rho$, where $ \rho
 \geq 0$. Upon doing so,  through 
the operational procedure corresponding to `integration by parts', we can
rewrite the expression
\begin{equation}
\chi^\rho\Upsilon(\chi) {\cal O}_\chi S(\chi)  = 
\Big (\sum_{j=0}^4 {U}_j(\chi^{\rho+9},\ldots,
\chi^{\rho-2};\rho) \Big ) S(\chi) + \partial_\chi \Big( T(\chi;S(\chi),\ldots,
S^{(4)}(\chi)) \Big ) = 0,
\end{equation}
involving the total derivative expression noted, and functions of the 
power expressions, $\chi^q$, where the exponent may be negative.

Since the physical solution must decay exponentially, along the asymptotic
directions ($\chi \rightarrow \pm \infty$), we can now integrate along a 
contour that sits on top of the real axis, except near the origin where
it deviates around it. Denote this contour by ${\cal C}$. Define the 
moments

\begin{equation}
\mu_\rho \equiv \int_{\cal C}d\chi \chi^\rho S(\chi).
\end{equation}
If $\rho \geq 0$, because of the analyticity of the solution, these moments
correspond to the Hamburger moments taken along the real axis:
$\mu_\rho = \int_{-\infty}^{+\infty} d\chi \chi^\rho \Psi(\chi)$, for $\rho \geq 0$. 

The moments will satisfy the moment equation (i.e. $p = \rho-2 \geq -2$)

\begin{eqnarray}
\mu_{p-3}\Big[ -{{3 E s_2 }\over b}p(4-4p-p^2+p^3)\Big] 
+\mu_{p-2} \Big[ -{{3 Es_2}\over b^2}p(-4-p+4p^2+p^3)\Big ] \cr
+\mu_{p-1}\Big[- {{E s_2}\over{b^3}}p(p+2)(6+12b^2c_2E + 7p +p^2-12b^5s_5)\Big] \cr
+\mu_p\Big[ {{6Es_2}\over{b^2}}(-2c_2E(4+5p+p^2)+s_5b^3(14+25p+8p^2))\Big] \cr
+\mu_{p+1}\Big[ {{2Es_2}\over{b^3}}(-2c_2E(12+8p+p^2)
+b^3s_5(168+169p+38p^2))\Big] \cr
+\mu_{p+2}\Big[{{12Es_2s_5}\over b}(42+30p+5p^2)\Big ] 
+\mu_{p+3}\Big[{{6Es_2}\over{b^3}}(18E^2s_2^2+bs_5(56+31p+4p^2))\Big] \cr
+\mu_{p+4}\Big[{{2Es_2}\over{b^4}}(162 E^2s_2^2 + b s_5(42+19p+2p^2))\Big ] 
+\mu_{p+5}\Big[ {{432 E^3s_2^3}\over {b^5}}\Big]
+\mu_{p+6}\Big[ {{324 E^3s_2^3}\over {b^6}}\Big] \cr
+\mu_{p+7}\Big[ {{144 E^3s_2^3}\over {b^7}}\Big]
+\mu_{p+8}\Big[ {{36 E^3s_2^3}\over {b^8}}\Big]
+\mu_{p+9}\Big[ {{4 E^3s_2^3}\over {b^9}}\Big] = 0,
\end{eqnarray}
where $p \geq -2$.

We note that for $p \geq 0$,   the
first nine Hamburger moments, $\{\mu_0,\ldots,\mu_8\}$ generate all 
of the other moments ($\mu_p$, $p \geq 9$) through the linear relation

\begin{equation}
\mu_p = \sum_{\ell = 0}^8{\tilde M}_{p,\ell}\mu_\ell,
\end{equation}
where the coefficients ${\tilde M}_{p,\ell}$ satisfy the moment equation (Eq.(35), for $p \geq 0$) with
respect to the $p$-index. In addition, they satisfy the initialization conditions ${\tilde M}_{\ell_1,\ell_2} = \delta_{\ell_1,\ell_2}$, for $0 \leq \ell_{1,2} \leq 8$.

If the moment equation is restricted to $p \geq 0$, then the associated 
differential equation would be of the form 
${\cal O}_\chi S(\chi) = {\cal D}(\chi)$, with ${\cal D}$ an unknown distribution supported at the origin (as discussed before). By also including $p = -1,-2$, we are imposing ${\cal D} = 0$. 
\vfil\break
For these
two values of the $p$ index, the moment equation becomes ($p = -1$)

\begin{eqnarray}
\mu_{0}\Big[ {{2Es_2}\over{b^3}}(-10c_2E + 37 b^3s_5)\Big]
+\mu_{1}\Big[{{204Es_2s_5}\over b}\Big ]
+\mu_{2}\Big[{{6Es_2}\over{b^3}}(18E^2s_2^2+29bs_5)\Big] \cr
+\mu_{3}\Big[{{2Es_2}\over{b^4}}(162 E^2s_2^2 + 25b s_5)\Big ]
+\mu_{4}\Big[ {{432 E^3s_2^3}\over {b^5}}\Big]  
+\mu_{5}\Big[ {{324 E^3s_2^3}\over {b^6}}\Big] 
+\mu_{6}\Big[ {{144 E^3s_2^3}\over {b^7}}\Big]\cr
+\mu_{7}\Big[ {{36 E^3s_2^3}\over {b^8}}\Big]
+\mu_{8}\Big[ {{4 E^3s_2^3}\over {b^9}}\Big] = 
\Sigma(\mu_{-1},\mu_{-2},\mu_{-4}),
\end{eqnarray}
where 
\begin{equation}
\Sigma(\mu_{-1},\mu_{-2},\mu_{-4})= \Big(18 b E s_2 s_5 \mu_{-1} -
{{E s_2}\over{b^3}}(12b^2c_2E-12b^5s_5)\mu_{-2} -{{18 E s_2 }\over b} \mu_{-4}\Big),
\end{equation}
and ($p = -2$)
\begin{eqnarray}
\mu_{0}\Big[{{24Es_2s_5}\over b}\Big ]
+\mu_{1}\Big[{{6Es_2}\over{b^3}}(18E^2s_2^2+10bs_5)\Big] 
+\mu_{2}\Big[{{2Es_2}\over{b^4}}(162 E^2s_2^2 + 12b s_5)\Big ]
+\mu_{3}\Big[ {{432 E^3s_2^3}\over {b^5}}\Big] \cr
+\mu_{4}\Big[ {{324 E^3s_2^3}\over {b^6}}\Big] 
+\mu_{5}\Big[ {{144 E^3s_2^3}\over {b^7}}\Big]
+\mu_{6}\Big[ {{36 E^3s_2^3}\over {b^8}}\Big]
+\mu_{7}\Big[ {{4 E^3s_2^3}\over {b^9}}\Big] = 
{2\over b} \Sigma(\mu_{-1},\mu_{-2},\mu_{-4}).
\end{eqnarray}
Therefore, we can use both relations to constrain the $\mu_{8}$ moment in 
terms of the moments $\{\mu_0,\mu_1,\ldots,\mu_7\}$. These in turn 
can be used with the moment equation (Eq.(35)), for $p \geq 0$, to
generate all of the other Hamburger moments. 
Thus, we have
\begin{equation}
\mu_8 = \sum_{\ell = 0}^7 {\tilde M}_{8,\ell}\mu_\ell,
\end{equation}
defined by subtracting Eq.(39) from Eq.(37) (after dividing the former by ${2\over b}$).

Incorporating this within the expression in Eq.(36),

\begin{equation}
\mu_p = \sum_{\ell = 0}^7{\tilde M}_{p,\ell}\mu_\ell +
{\tilde M}_{p,8} \sum_{\ell = 0}^7 {\tilde M}_{8,\ell}\mu_\ell,
\end{equation}
yields
\begin{equation}
\mu_p = \sum_{\ell = 0}^7 M_{p,\ell}\mu_\ell,
\end{equation}
where
\begin{equation}
M_{p,\ell} = \cases{ {\tilde M}_{8,\ell},\ p = 8 \cr
	      {\tilde M}_{p,\ell} + {\tilde M}_{p,8} {\tilde M}_{8,\ell},\ p \geq 9 \cr},
\end{equation}
and $M_{\ell_1,\ell_2} = \delta_{\ell_1,\ell_2}$, for $0 \leq \ell_1,\ell_2 \leq 7$.
This is the moment generating relation required in implementing EMM.

An additional important point is specifying the choice of normalization.
Since we are only dealing with Hamburger moments, some of the odd order moments
may be negative. However, the even order moments are positive, and by working
with the first 8 of them 

\begin{equation}
u_\ell \equiv \mu_{2\ell},
\end{equation}
we can impose the normalization

\begin{equation}
\sum_{\ell = 0}^7 u_\ell = 1.
\end{equation}
Working with the positive quantities, $u_\ell$, makes the implementation of the 
linear programming algorithm more efficient (since one is working within a
bounded polytope corresponding to the unit hypercube, $[0,1]^7$).

Thus, we take

\begin{equation}
u_\ell = \mu_{2\ell} = \sum_{\ell_v}^7 M_{2\ell,\ell_v} \mu_{\ell_v},
\end{equation}
for $0 \leq \ell \leq 7$. The inverse of this relation is denoted by
\begin{equation}
 \mu_{\ell} = \sum_{\ell_v=0}^7 N_{\ell,\ell_v} u_{\ell_v},
\end{equation}
and substituted in Eq.(42) producing

\begin{equation}
\mu_p = \sum_{\ell,\ell_v=0}^7M_{p,\ell} N_{\ell,\ell_v} u_{\ell_v},
\end{equation}
or
\begin{equation}
\mu_p = \sum_{\ell = 0}^7 {\cal M}_{p,\ell} u_\ell,
\end{equation}
for $ p \geq 0$, where ${\cal M}_{p,\ell} = \sum_{\ell_v=0}^7M_{p,\ell_v} N_{\ell_v,\ell}$. 

Finally, the normalization condition is incorporated
($u_0 = 1 -\sum_{\ell = 1}^7 u_\ell$), producing

\begin{equation}
\mu_p = \sum_{\ell = 0}^7 {\hat{\cal M}}_{p,\ell} {\hat u}_\ell,
\end{equation}
where
\begin{equation}
 {\hat{\cal M}}_{p,\ell} = \cases { {\cal M}_{0,\ell},\ p = 0 \cr
			{\cal M}_{p,\ell} - {\cal M}_{0,\ell},\ p \geq 1\cr },
\end{equation}
and
\begin{equation}
{\hat u}_\ell = \cases{ 1, \ell = 0 \cr
	                u_\ell, 1 \leq \ell \leq 7\cr }.
\end{equation}
\subsection{ Numerical Results}

The numerical implementation of the EMM procedure is given in Table I. 
We only investigate the lowest energy state. The
parameter $P_{max}$ denotes the maximum moment order generated, $\{ \mu_p |
0 \leq p \leq P_{max}\}$. We see that as the rotation angle increases, the
tightness of the bounds decreases, as expected. Furthermore, for $\theta >
{\pi\over {10}}$, no EMM solution is generated, consistent with the
fact that the bound state becomes exponentially unbounded, in the corresponding direction
in the complex $x$ plane.

\begin{table}
\caption {Bounds for the Ground State Energy of the $-ix^3$ Potential
Using Eq.( ) ($P_{max} = 40$)} 
\begin{center}
\begin{tabular}{cccccl}
 \multicolumn{1}{c}{$\theta$}
& \multicolumn{1}{c}{$E_{L;0}$} & \multicolumn{1}{c}{$E_{U;0}
$}\\ \hline
$0^*$ & 1.15619  &        1.15645 \\
.01 & 1.15617 &        1.15652 \\
.05 & 1.15510 &        1.15761 \\
.10 & 1.14201 &   1.17298 \\
.15 & 1.0 &   1.3 \\
.20 & .5 & 8.0 \\
\end{tabular}
\end{center}
\noindent{*Using Stieltjes EMM formulation by
Handy(2001)}
\end{table}

\vfil\break
\section {Acknowledgments}
This work was supported through a grant from the National Science Foundation 
(HRD 9632844) through the Center for Theoretical Studies of Physical Systems
(CTSPS). The authors are appreciative of relevant comments by D. Bessis.

\vfil\break
\section {References}
\noindent Bender C M and Boettcher S 1998 Phys. Rev. Lett. {\bf 80} 5243

\noindent Bender C M, Boettcher S, and Meisinger P N 1999, J. Math. Phys. {\bf 40} 2201

\noindent Bender C M, Boettcher S, Jones H F and Savage V M 1999 J. Phys. A:
Math. Gen. {\bf 32} 1

\noindent Bender C M, Boettcher S and Savage V M 2000 J. Math. Phys. {\bf 41} 6381

\noindent Bender C M and Wang Q 2001 J. Phys. A: Math. Gen.  

\noindent Bender C M and Orszag S A,   {\it Advanced Mathematical
Methods for
Scientists and Engineers} (New York: McGraw Hill 1978).

\noindent Caliceti E 2000 J. Phys. A: Math. Gen. {\bf 33} 3753

\noindent Chvatal V 1983 {\it Linear Programming} (Freeman, New York).

\noindent Delabaere E and Pham F 1998 Phys. Lett. {\bf A250} 25

\noindent Delabaere E and Trinh D T 2000 J. Phys. A: Math. Gen. {\bf 33} 8771

\noindent Handy C R 1987a Phys. Rev. A {\bf 36}, 4411 

\noindent Handy C R 1987b Phys. Lett. A {\bf 124}, 308 

\noindent Handy C R 2001 (LANL math-ph/0104036, to appear in J. Phys. A)

\noindent Handy C R and  Bessis D 1985 Phys. Rev. Lett. {\bf 55}, 931 

\noindent Handy C R, Bessis D, and Morley T D 1988a Phys. Rev. A {\bf 37}, 4557 

\noindent Handy C R, Bessis D, Sigismondi G, and Morley T D 1988b Phys. Rev. Lett
{\bf 60}, 253 

\noindent Handy C R, Khan D,  Wang Xiao-Qian, and Tymczak C J 2001 (LANL math-ph/0104037, submitted to J. Phys. A)

\noindent Handy C R, Luo L, Mantica G, and Msezane A 1988c Phys. Rev. A {\bf 38}, 490

\noindent Handy C R 2001, CAU preprint (to appear in  J. Phys. A)

\noindent Handy C R and Msezane A Z 2001, CAU preprint

\noindent Handy C R, Trallero-Giner C and Rodriguez A Z (2001) CAU preprint (submitted to J. Phys. A)

\noindent Levai G and Znojil M 2000 J. Phys. A: Math. Gen. {\bf 33} 7165

\noindent Mezincescu G A 2000 J. Phys. A: Math. Gen. {\bf 33} 4911

\noindent Mezincescu G A 2001 J. Phys. A: Math. Gen.

\noindent Shohat J A and  Tamarkin J D, {\it The Problem of Moments}
(American Mathematical Society, Providence, RI, 1963).

\noindent Shin K C 2000 Preprint math-ph/0007006 (J. Math. Phys, under press)

\noindent Znojil M 2000 J. Phys. A: Math. Gen. {\bf 33} 6825

\end{document}